# Comprehensive Resource Measurement and Analysis for HPC Systems

Charng-Da Lu

## 1. INTRODUCTION
High-performance computing (HPC) systems are a complex combination of software, processors, memory, networks, and storage systems characterized by frequent disruptive technological advances. Anomalous behavior has to be manually diagnosed and remedied with incomplete and sparse data. It also has been effort-intensive for users to assess the effectiveness with which they are using the available resources. The data available for system level analyses appear from multiple sources and in disparate formats (from Linux "sysstat" [2] and accounting to scheduler/kernel logs). Sysstat does not resolve its measurements by job so that job-oriented analyses require individual measurements. There are many user-oriented performance instrumentation and profiling tools but they require extensive system knowledge, code changes and recompilation, and thus are not widely used. In this environment, system managers, users and sponsors find it difficult if not impossible to know if the systems are being utilized effectively at any level or even if all subcomponents are functioning properly. HPC centers and their users have to some extent been "flying blind", without a clear understanding of system behavior. Del Vento and *et al*. [4] adopted a similar approach to tackling inefficient HPC resource utilization. With detailed system-wide monitoring and user collaboration, they were able to improve CPU usage (e.g. correct process affinity/CPU binding) and troubleshoot thorny issues (e.g. memory leak.) However, their work is based on proprietary systems (IBM POWER/AIX) while ours is designed for open-source software based HPC clusters.

TACC_Stats is a newly developed job-oriented and logically structured version of the conventional Linux "sysstat/sar" system-wide performance monitor. In addition to the data traditionally collected by sysstats/sar, TACC_Stats records hardware performance counter values, parallel file-system metrics, and high-speed interconnect usage. The core component is a collector executed on every compute nodes, both at the beginning/end of job (via batch scheduler prolog and epilog scripts) and at periodic intervals (via cron). Like the "sar" (System Activity Reporter) program of the sysstat package, the performance collection takes place in the background, incurs very low overhead, and requires no user intervention.

## 2. RESOURCE USAGE MEASUREMENT
The Linux sysstat package is a comprehensive collection of performance monitoring utilities, each of which reports resource statistics of specific components of a system in its own format. TACC_Stats enhances sysstat/sar for HPC environment in many ways. It is a single executable binary which implements a superset of the measurement functions of sysstat and outputs in a unified, consistent, and self-describing plain-text format. It is job aware: resource usage data at the core and node levels are tagged with job-ids to enable offline job-by-job profile analysis. It supports newer Linux counters and hardware devices. Its source code is also highly modular and can be easily extended to gather new kinds of performance metrics.

Currently TACC_Stats can gather core-level CPU usage (user time, system time, idle, etc), socket-level memory usage (free, used, cached, etc), swapping/paging activities, system load and process statistics, network and block device counters, interprocess communications (SysV IPC), software/hardware interrupt request (IRQ) count, filesystems usage (NFS, Lustre, Panasas), interconnect fabric traffic (InfiniBand and Myrinet), and CPU hardware performance counters.

TACC_Stats utilizes CPU performance counters as follows. At the beginning of job, TACC_Stats is invoked by the batch scheduler prolog to reprogram performance counters to record a fixed set of events: On AMD Opteron, the events are FLOPS, memory accesses, data cache fills, SMP/NUMA traffic. On Intel Nehalem/Westmere, FLOPS, SMP/NUMA traffic, and L1 data cache hits. TACC_Stats then records this data at intervals (currently set at 10 minutes) during execution. To avoid interfering with user's profiling and instrumentation activities, TACC_Stats only reads values from performance counter registers without reprogramming them.

TACC_Stats may be downloaded [3] and is readily installed on many versions of Linux. It has, for example, been installed on the Edge cluster at the State University of New York at Buffalo.

## 3. CASE STUDIES
The initial case studies reported here were done using data from the Ranger supercomputer at the Texas Advanced Computing Center (TACC). It is a Linux cluster comprising of 3936 nodes, each of which has four 2.3GHz AMD Opteron quad-core processors (16 cores in total) and 32 GB of memory. The filesystem is Lustre and the interconnect is InfiniBand. TACC_Stats has been deployed on Ranger for about a year. It generates a raw data file of 0.5 MB per node per day and collectively 60 GB (uncompressed) or 20 GB (compressed) for the entire cluster per month.We analyzed TACC_Stats data collected during January 5 - February 10, 2012. We ingested both the raw TACC_Stats output and job accounting information into MySQL database. There are 50338 jobs in total. Excluding small jobs (#Nodes * Wall Clock Time less than one node-hour) or jobs submitted to non-production queues, we are left with 24997 jobs. We will look for inefficient jobs in this job pool in the following case studies.

### 3.1 CPU Idle Fraction
About 9.7% of the aggregated CPU time of the 24997 jobs is idle. Although high idle fraction is a proxy for low computational intensity, a user could have valid reasons to use only a fraction of available CPU cycles, e.g. the application is memory hungry but computationally light. Note, however, that it is also possible that a job with a low CPU idle fraction may not be executing with maximum efficiency due to bottlenecking on memory access or not taking advantage of "vector" arithmetic instructions. On

Ranger, a user can control a job's process affinity by specifying the "wayness", i.e. the number of MPI processes to be launched on each node. Figure 1 shows the CPU and memory profile of the jobs. Seventy-eight percent of jobs had good CPU usage (≤10% CPU idle time) and the 32GB node memory is sufficient for most jobs. The salient vertical stripes of red dots at CPU idle fraction of 0.5, 0.75, 0.875, and 0.9375 are jobs which only need partial CPU resources (8, 4, 2, and 1-way). There are also many non-16-way jobs which use almost all cores and this is probably due to their multithread-ness. Jobs appearing at the upper-right corner of Figure 1 are those which are the most wasteful of CPU resource. To better pinpoint these "bad" jobs, we calculated a derived metric "CPU idle fraction×Unused memory fraction." and plotted this data. We then identified 47 jobs with this metric being greater than 0.9.

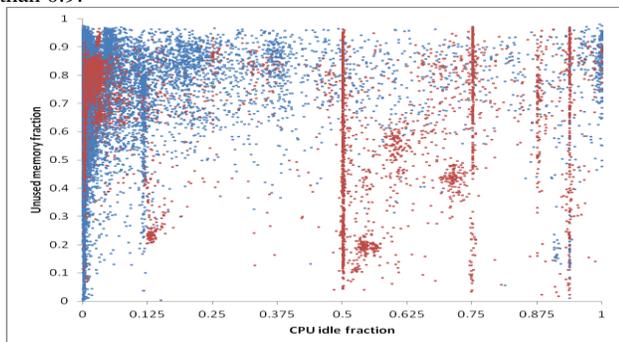

Figure 1: CPU and memory usage of jobs. Blue dots are 16-way jobs and red ones are non-16-way.

Among them, 29 are from three users and their fields of science are material science and biology. The choice of 0.9 as the threshold for "bad" is arbitrary and we will investigate benefits of choosing a different threshold.

### 3.2 Memory Access Imbalance

The memory organization of a Ranger node is based on the Non-Uniform Memory Access (NUMA) architecture: Each CPU has its own local memory and a CPU can access other's memory via the on-board fabric. On Ranger, a user can customize the memory policy (i.e. memory allocation strategy) for their jobs, and an optimal setting should result in a homogeneous memory access pattern across all CPUs and hence better performance. Figure 2 shows the jobs' sustained memory bandwidth and the imbalance-ness (measured by **C**oefficient **o**f **V**ariation=StdDev/Mean) among the four CPUs in each node. Understandably non-16-way jobs tend to have higher CoVs since they don't even need to use all CPUs. If we focus on jobs which are 16-way, memory access rate greater than 1 GB/s, and CoV higher than 1, there are 68 jobs. These "bad" jobs have thin "shape" (1-2 nodes but very long running time) and 82% of them are from a single user (material science).

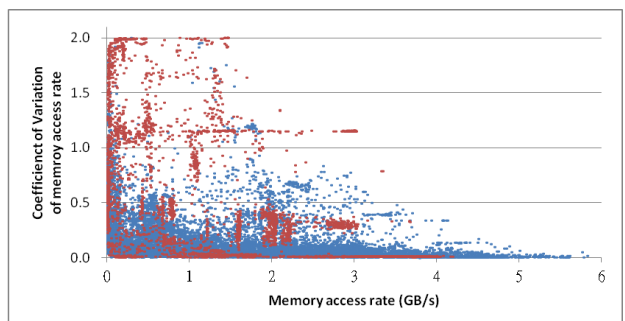

Figure 2: Sustained memory bandwidth and access imbalance of jobs. Blue dots are 16-way jobs and red ones are non-16-way.

### 4. FUTURE WORK

We are contacting the job owners for the high idle fraction and high NUMA access fractions to see if they are interested in collaboration to understand the cause or causes for these poor performance metrics. We will be pursuing several other job-user oriented case studies such as effectiveness of I/O use and detection of memory leaks. We are also assessing various technologies (e.g. NoSQL) to quickly process, store, and query massive TACC_Stats data. We plan to leverage the XDMOD [5] project tools to develop easy-to-use exploratory and diagnosis tools for TACC_Stats data for both systems administrators and users. TACC_Stats [1] will be extended and modified as dictated by experience. We will also explore the possibility of using the extensive data available in developing predictive models for "Application Kernels" [5], a set of well-optimized and lightweight codes that form the core of many HPC applications and workloads Finally, we are planning to extend the FDiag [6] log analysis framework, which currently diagnoses system faults and failures to do more detailed studies of system behaviors such as job scheduler policies.

### REFERENCES

[1] J. Hammond. "TACC_stats: I/O performance monitoring for the intransigent" In *2011 Workshop for Interfaces and Architectures for Scientific Data Storage (IASDS 2011)*

[2] http://sebastien.godard.pagesperso-orange.fr

[3] http://github.com/TACCProjects/tacc_stats

[4] D. Del Vento, T. Engel, S. Ghosh, D. Hart, R. Kelly, S. Liu, R. Valent. "System-level monitoring of floating-point performance to improve effective system utilization." In *2011 International Conference for High Performance Computing, Networking, Storage and Analysis* (*SC11*)

[5] T. Furlani and *et al*. "Performance metrics and auditing framework using applications kernels for high performance computer systems." In *2011 TeraGrid Conference* (*TG11*). http://xdmod.ccr.buffalo.edu

[6] E. Chuah, G. Lee, WC. Tjhi, SH. Kuo, T. Hung, J. Hammond, T. Minyard, J.C. Browne. "Establishing hypothesis for recurrent system failures from cluster log files." In *Proceedings of the IEEE International Conference on Dependable, Autonomic and Secure Computing, 2011*